\documentclass[prl,twocolumn,superscriptaddress,showpacs,amsmath,amssymb]{revtex4}

\usepackage{graphicx}
\usepackage{dcolumn}
\usepackage{bm}
\usepackage{color}

\begin{document}


\title{Edge States and Quantum Hall Effect in Graphene under a Modulated Magnetic Field}

\author{Lei Xu}
\affiliation{National Laboratory of Solid State Microstructures and
Department of Physics, Nanjing University, Nanjing 210093, China}

\author{Jin An}
\email{anjin@nju.edu.cn}
\affiliation{National Laboratory of Solid
State Microstructures and Department of Physics, Nanjing University,
Nanjing 210093, China}

\author{Chang-De Gong}
\affiliation{Center for Statistical and Theoretical Condensed Matter
Physics, and Department of Physics, Zhejiang Normal University,
Jinhua 321004, China}

\affiliation{National Laboratory of Solid State Microstructures and
Department of Physics, Nanjing University, Nanjing 210093, China}

\date{\today}

\begin{abstract}
Graphene properties can be manipulated by a periodic potential.
Based on the tight-binding model, we study graphene under a
one-dimensional (1D) modulated magnetic field which contains both a
uniform and a staggered component. The chiral current-carrying edge
states generated at the interfaces where the staggered component
changes direction, lead to an unusual integer quantum Hall effect
(QHE) in graphene, which can be observed experimentally by a
standard four-terminal Hall measurement. When Zeeman spin splitting
is considered, a novel state is predicted where the electron edge
currents with opposite polarization propagate in the opposite
directions at one sample boundary, whereas propagate in the same
directions at the other sample boundary.

\end{abstract}

\maketitle

Recently, graphene materials have received extensive theoretical and
experimental studies\cite{Castro Neto2009}. The most important
physical properties of graphene are governed by the underlying
chiral Dirac fermions\cite{Zhou2006,Li2007}. These Dirac fermions
under a uniform magnetic field (UMF) give rise to the well-known
anomalous QHE\cite{Novoselov2005,Zhang2005}, which has been
experimentally verified and is now believed to be a unique feature
to characterize graphene. Spin QHE was also predicted in graphene,
where electron edge current with the opposite spin polarization
couterpropagates due to the spin-orbit interaction\cite{Kane2005} or
the Zeeman spin splitting of the zeroth Landau level
(LL)\cite{Abanin2006}. On the other hand, the experimental
manipulation of the electronic structure of graphene has potential
application in graphene electronics or
spintronics\cite{Rycerz2007,Son2006}. One method to manipulate the
physical properties of graphene is by applying periodic
electronic\cite{Park2008,Park2009} or magnetic
potentials\cite{DellAnna2009,DellAnna2009-2}, which can be realized
now by making use of
substrate\cite{Vazquez2008,Martoccia2008,Sutter2008,Pletikosic2009}
or controlled adatom deposition\cite{Meyer2008}.

\begin{figure}
\scalebox{0.5}[0.5]{\includegraphics[91,427][557,738]{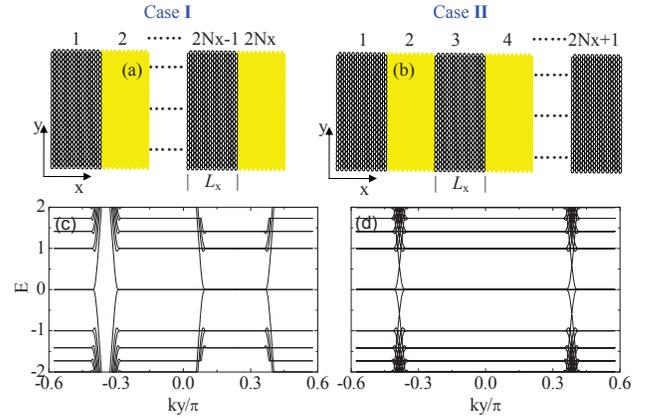}}
\caption{\label{fig:fig1}(color online). Top panels: Schematic
illustration of a rectangular graphene sample under a SMF, where two
uniform fields which are the same in magnitudes but opposite in
directions alternate every $Lx$ distance. Case \textbf{I} in (a)
contains $2N_x$ alternating regions whereas Case \textbf{II} in (b)
contains $2N_x+1$ alternating regions, with $N_x$ an integer. Bottom
panels: Electron energy bands of graphene under a periodically SMF
with $L_x=426$ nm, and $B_S=24$ T for (c), $B_S=40$ T for (d).
Energy is measured in unit of $\hbar v_F/l_B$ with $l_B$ the
magnetic length. For the parameters chosen, $l_B$ is determined to
be $l_B=3.7$ nm for (c), $l_B=2.9$ nm for (d), respectively.}
\end{figure}

Here, we report the investigation of the effect on graphene QHE of a
1D staggered magnetic field (SMF), which is schematically shown in
the top panels of Fig.~\ref{fig:fig1}.  The 1D SMF can be achieved
in experiments by applying an array of ferromagnetic stripes with
alternative magnetization on the top of a graphene layer. It is
found that the edge states created by 1D SMF lead to a nontrivial
robust integer QHE in graphene. In a standard four-terminal Hall
measurement, when varying the magnitude of the UMF, graphene can
undergo a transition from a state with unusual quantized Hall
conductance to one without Hall effect. Furthermore, the Zeeman spin
splitting of the zeroth LL of graphene gives rise to a novel state
where spin-up and spin-down edge currents have the opposite
chirality at one sample boundary but have the same chirality at the
other sample boundary.

\begin{figure}
\scalebox{0.5}[0.5]{\includegraphics[89,275][560,736]{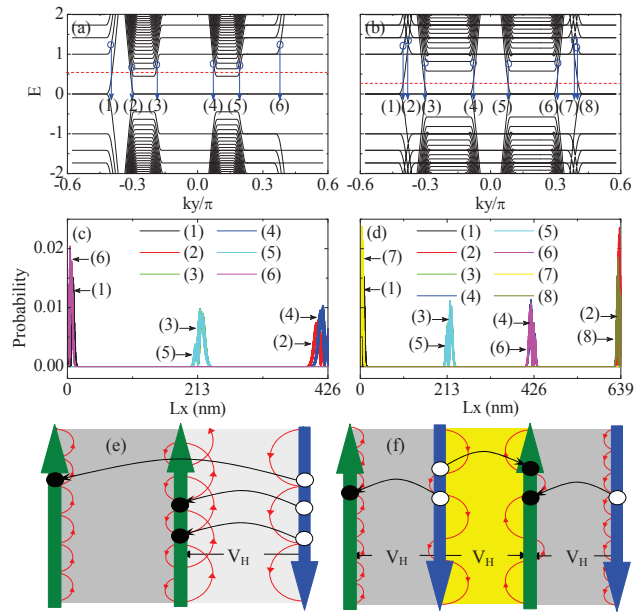}}
\caption{\label{fig:fig2}(color online). Top panels: Electron energy
spectrum of graphene in the presence of both a UMF and a SMF for
$N_x=1$ and $L_x=213$ nm. (a) belongs to Case \textbf{I} with
parameters $B_U=24$ T, $B_S=16$ T. (b) belongs to Case \textbf{II}
with parameters $B_U=16$ T, $B_S=32$ T. Here energy is scaled in the
same way to Fig.~\ref{fig:fig1} except the magnetic length $l_B$ is
replaced by that for the high-field regions. The red dashed lines
represent the positions of the chemical potentials. Middle panels:
The corresponding electron probability densities for the
representative states indicated in the top panels by the open
circles. Bottom panels: Schematic diagrams of the corresponding
electron edge currents (green and blue arrows) and electrons'
classical skipping orbits. $V_H$ is the Hall voltage. Gray (light
gray, yellow) solid rectangle represents high-(low-) field regions.
Black and white solid circles represent particle-hole excitations in
the Laughlin's thought experiment.}
\end{figure}

We start from the tight-binding model on a honeycomb lattice in a
perpendicular nonuniform magnetic field described by the
Hamiltonian,
\begin{equation}
H=-t\sum_{<ij>}e^{i\int_i^j\textbf{A}\cdot d\textbf{l}}c_{i}^\dag
c_{j}+ \texttt{H.c.}, \label{eq:one}
\end{equation}
where $t$ is the hopping integral, the operator $c_i^\dag$ ($c_i$)
creates (annihilates) an electron at site $i$, and $<ij>$ denotes
nearest-neighbor pairs of sites. \textbf{A} is the gauge potential
for the nonuniform magnetic field. The Zeeman spin splitting is
neglected now for simplicity and will be discussed later. We
distinguish here Case \textbf{I} in Fig.~\ref{fig:fig1}(a) from Case
\textbf{II} in Fig.~\ref{fig:fig1}(b) because graphene in Case
\textbf{I} has no QHE unless the magnetic fields of the two
alternating areas have the same directions, i.e. $B_U>B_S$ where
$B_U$ and $B_S$ are the magnitudes of the UMF and SMF respectively.
The reason for this will be explained later when we discuss
Fig.~\ref{fig:fig2}.

The LLs of graphene\cite{McClure1956} can be expressed as
$E_n=\pm\hbar v_F\sqrt{|n|}/l_B$ with $l_B=\sqrt{\hbar c/2e B}$ the
magnetic length, $v_F=3at/2\hbar$ the Fermi velocity, and
$n=0,1,2,...$ the LL index. Here $a$ is graphene lattice constant.
The physical picture at large $L_x$ order of $L_x>>l_B$ is found to
be quite different from that at smaller $L_x$ case with $L_x\sim
l_B$, where only a few chains are contained in each alternating
area. In the latter case, though Dirac cone structure is preserved,
more and more Dirac points are created as increasing $L_x$\cite{Xu},
and finally the LLs of graphene appear. In Figs.~\ref{fig:fig1}(c)
and \ref{fig:fig1}(d) we show in the absence of the UMF the electron
energy bands of graphene under a periodic SMF with $k_x=0$. For the
SMF with a period $2L_x$, we have chosen correspondingly a periodic
gauge $\textbf{A}=\textbf{A}_S=(0,B_S |x|,0)$ for $|x|<L_x$. When
energy is set by $\hbar v_F/l_B$, the $\sqrt{n}$ spacing of the LLs
of graphene can be seen clearly. What's remarkable is that for large
$L_x$, the dense Dirac points are emerged into the zeroth LL of
graphene\cite{Xu}. Compared with the LLs of graphene in a UMF, which
is dispersionless, the energy bands have non-flat regions where
energy disperse with $k_y$, indicating the presence of edge states.
These new edge states which were called magnetic edge
states\cite{Reijniers2001,Sim1998} are actually generated by the SMF
and right located at the interfaces where the SMF changes direction,
providing the gapless excitations for even an infinite graphene.

To further clarify the physical properties of these edge states and
its consequences, we study the QHE of such a system, i.e., the
graphene samples in the presence of both a UMF and a SMF. In the
following cases, open boundary condition is applied in the $x$
direction and periodic boundary condition in the $y$ direction, and
the Landau gauge $\textbf{A}_U=(0,B_Ux,0)$ is adopted for the UMF
and $\textbf{A}=\textbf{A}_S+\textbf{A}_U$. The graphene energy
spectrum for Case \textbf{I} and Case \textbf{II} are calculated and
shown in Figs.~\ref{fig:fig2}(a) and \ref{fig:fig2}(b),
respectively. With application of a UMF, the graphene samples can be
divided into two groups of regions, where one is the high-field
group of regions with the magnetic field $B_U+B_S$, the other is the
low-field group of regions with the magnetic field $B_U-B_S$.
Correspondingly, the bulk excitations of the graphene system have
two groups of LLs, which are reflected in electron energy bands by
the two groups of the flat regions in Figs.~\ref{fig:fig2}(a) and
\ref{fig:fig2}(b). Interestingly, when the ratio
$(B_U-B_S)/(B_U+B_S)=p/q$, where $p$ and $q$ are two coprime
integers, a series of LLs will be doubly degenerate, which may cause
some interesting phenomena. In particular, we actually have
$(B_U-B_S)/(B_U+B_S)=1/5$ and $-1/3$ for the parameters in
Figs.~\ref{fig:fig2}(a) and \ref{fig:fig2}(b), respectively,
resulting in the doubly degeneracy of all the LLs in the high-field
regions.

Now we focus our attention on the edge states. In
Figs.~\ref{fig:fig2}(c) and \ref{fig:fig2}(d), electron probability
densities for representative edge states are shown. Except the
conventional edge states located at the sample boundaries [see
(1),(2),(4),(6) in Fig.~\ref{fig:fig2}(c), and (1),(2),(7),(8) in
Fig.~\ref{fig:fig2}(d)], new edge states are generated and right
localized at the interfaces where the SMF changes direction [see
(3),(5) in Fig.~\ref{fig:fig2}(c), and (3)-(6) in
Fig.~\ref{fig:fig2}(d)]. Clearly, these edge states are also chiral
current-carrying states\cite{s-Park2008,Oroszlany2008,Ghosh2008},
whose flowing directions can be easily determined by the slopes of
the bands where they are located. All the edge currents are
schematically shown in the bottom panels of Fig.~\ref{fig:fig2}.
Also indicated in these panels are the classical orbits of
electrons\cite{Muller1992}, which are composed of arcs with the
length scale approximately equal to the magnetic length $l_B$. These
classical orbits give a simple physical interpretation of the reason
why the edge currents flow in the shown directions.


\begin{figure}
\scalebox{0.5}[0.5]{\includegraphics[86,581][524,738]{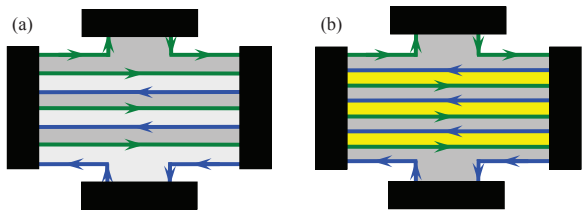}}
\caption{\label{fig:fig3}(color online). The four-terminal Hall bar
geometries for (a) Case \textbf{I} and (b) Case \textbf{II} with
$N_x=3$. The black solid rectangles represent contacts and the lines
with arrows represent the edge currents. }
\end{figure}

Let us now consider the corresponding four-terminal Hall
measurements. Assuming that contacts are reflectionless, all edge
currents coming from the same contact share the same voltage with
that contact, so that all the ``\,up-flowing\," edge currents have
the same voltage and so do all the ``\,down-flowing\," edge
currents. The voltage difference between the edge currents with
opposite directions is equivalent to the Hall voltage $V_H$ between
the sample boundaries. To obtain the Hall conductance in these two
situations in Figs.~\ref{fig:fig2}(e)-(f), we follow Laughlin's
argument\cite{Laughlin1981} and imagine rolling up the graphene
ribbon along the $y$ direction to make a graphene cylinder which is
then threaded by a magnetic flux. When varying adiabatically the
magnetic flux through the graphene cylinder by a flux quantum
$\phi_0$, particle-hole excitations will be generated both at the
sample boundaries and the interfaces where the SMF changes
direction. For an illustration, we show schematically in
Figs.~\ref{fig:fig2}(e)-(f) these excitations for the chemical
potentials indicated by the red lines in
Figs.~\ref{fig:fig2}(a)-(b). The increase in energy due to electrons
transfer between the Hall voltage is $\delta E=3eV_H$ for both
Figs.~\ref{fig:fig2}(e) and \ref{fig:fig2}(f). Generally, if the
system has such a Fermi energy that the $M$th LL of the high-field
regions and the $N$th LL of the Low-field regions are just
completely filled, detailed analysis of the edge states in the
energy bands shows that the increase in energy $\delta E$ can be
given by $\delta E=P_NeV_H$ for Case \textbf{I} in
Fig.~\ref{fig:fig2}(e), and $\delta E=(P_N+2P_M)eV_H$ for Case II in
Fig.~\ref{fig:fig2}(f), respectively, where $P_M=2M+1$ and
$P_N=2N+1$. Hence, from $I=c\delta E/\delta\phi$, the Hall
conductance is given by $\sigma_{xy}=P_Ne^2/h$ for Case \textbf{I}
in Fig.~\ref{fig:fig2}(e), which has nothing to do with the
high-field LLs, and $\sigma_{xy}=(P_N+2P_M)e^2/h$ for Case II in
Fig.~\ref{fig:fig2}(f).

We note that in order to have the mentioned QHE for Case
\textbf{I}(\textbf{II}), the magnetic field for the high-field and
low-field regions must have the same (opposite) direction, i.e.,
$B_U>B_S$ ($B_U<B_S$). For Case \textbf{I} in
Fig.~\ref{fig:fig2}(e), by varying the UMF so that the magnetic
field for the low-field region is reversed, i.e., $B_U<B_S$, the
electron edge currents at the right edge and the interface will be
reversed too, resulting in the same voltage shared by the two sample
boundaries and thus a zero Hall voltage. For Case \textbf{II} in
Fig.~\ref{fig:fig2}(f), however, the reversal of the magnetic field
in the low-field region for $B_U>B_S$ only leads to the reversal of
edge currents at the two interfaces with that at the two sample
boundaries unchanged, giving rise to another quantized Hall
conductance $\sigma_{xy}=P_Ne^2/h$. Therefore, for both cases,
$B_U=B_S$ is a critical value for graphene QHE under a SMF. Another
important point we should remark is that for Case \textbf{II} in
Fig.~\ref{fig:fig2}(e), even in the absence of a UMF, graphene under
a purely SMF shows a quantized Hall conductance
$\sigma_{xy}=3P_Ne^2/h$, where $P_N=P_M$. This can be comparable
with the Haldane model introduced in
Ref.~[\onlinecite{Haldane1988}], where there exists a nonzero
quantized Hall conductance in the absence of an external UMF. All
these peculiar behaviors are believed to have great application in
graphene manipulation.

It is straightforward to generalize the scheme used above to more
general and interesting cases for $N_x>1$. Our detailed calculation
confirm the existence of the two groups of the LLs for both Cases,
which are the bulk excitations of graphene under the modulated
magnetic field and are represented by the flat bands in the electron
spectrum. Except the LLs and conventional edge states at the sample
boundaries, there exist many current-carrying edge states localized
at the interfaces where the SMF changes direction. In
Fig.~\ref{fig:fig3}, we take $N_x=3$ for example, and show
schematically the corresponding four-terminal measurements, where
$N_x$ now has the meaning of the number of the low-field regions.
Detailed analysis of these edge states from the spectrum and similar
argument lead to the Hall conductance as follows:

\noindent when $B_U>B_S$, for both Case \textbf{I} and \textbf{II},
\begin{equation}
\sigma_{xy}^{\text{\textbf{I}}}=\sigma_{xy}^{\text{\textbf{II}}}=\pm
\frac{e^2}{h}[N_xP_N-(N_x-1)P_M], \label{eq:2}
\end{equation}
when $B_U<B_S$, for Case \textbf{II},
\begin{equation}
\sigma_{xy}^{\text{\textbf{II}}}=\pm
\frac{e^2}{h}[N_xP_N+(N_x+1)P_M], \label{eq:3}
\end{equation}
whereas for Case \textbf{I}, there is no Hall voltage. Here the
``$\pm$" symbol represents the particle-hole symmetry. If $N_x=1$,
the previous results are recovered. The result Eq.~(\ref{eq:3})
seems trivial since it can be seen as the conductance sum of all the
contributions from each distinct region. The result
Eq.~(\ref{eq:2}), however, is highly nontrivial since it can not be
seen as the naive subtraction of the contribution of the high-field
regions from that of the low-field regions. When the chemical
potential is within a gap between the LLs, while the Hall
conductance is quantized, the resistance of system should be
vanishingly small because only the edge states carry current so that
the backscattering is strongly suppressed. Due to the gauge
invariance of Laughlin's argument used here, the results we obtained
should be robust against weak disorder and interaction. On the other
hand, by varying the magnitude of the UMF, we come to a similar
conclusion to the cases with $N_x=1$ that, the graphene will undergo
a transition, from one state with the quantized Hall conductance
Eq.~(\ref{eq:2}) to one without Hall effect for Case \textbf{I} in
Fig.~\ref{fig:fig3}(a), whereas from one state with the Hall
conductance Eq.~(\ref{eq:3}) to one with the Hall conductance
Eq.~(\ref{eq:2}) for Case \textbf{II} in Fig.~\ref{fig:fig3}(b).

\begin{figure}
\scalebox{0.5}[0.5]{\includegraphics[83,379][561,734]{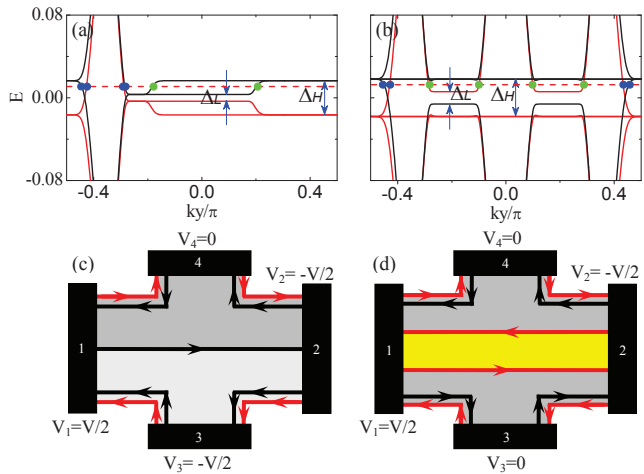}}
\caption{\label{fig:fig4}(color online). Top panels: Electron energy
spectrum of graphene in the presence of Zeeman splitting for (a)
Case \textbf{I} and (b) Case \textbf{II} with $N_x=1$. The Zeeman
energy is chosen to be one tenth of the largest LL spacing of the
high-field regions. The red dashed lines represent the positions of
the chemical potentials. The blue filled circles indicate the edge
states at the sample boundaries whereas the green filled circles
indicate the newly generated edge states localized at the interfaces
where the SMF changes direction. Bottom panels: The corresponding
four-terminal Hall bar geometries, where the red and black lines
with arrows represent the spin-up and spin-down edge currents,
respectively.}
\end{figure}

Now we turn to explore the spin effect in graphene QHE in the
presence of Zeeman spin splitting. The energy spectrum is shown in
Figs.~\ref{fig:fig4}(a)-(b). Two spin gaps $\Delta_H$ and
$\Delta_L$, which are corresponding to the Zeeman splitting in the
high-field and low-field regions respectively, are opened. When the
chemical potential lies within the interval
$-\Delta_L/2<\mu<\Delta_L/2$, there is no edge current at the
interfaces where the SMF changes direction, as well as two edge
states at both sample boundaries, which have opposite spin
polarizations and propagate in opposite directions\cite{Abanin2006}.

When the chemical potential lies within the interval
$\Delta_L/2<|\mu|<\Delta_H/2$, fully spin-polarized edge currents
appear at the interfaces where the SMF changes direction (See the
four-terminal Hall measurements shown in
Figs.~\ref{fig:fig4}(c)-(d)), which is spin-down for $B_U>B_S$, and
spin-up for $B_U<B_S$. Remarkably, for Case \textbf{I} in
Fig.~\ref{fig:fig4}(c), a novel state occurs where at one sample
boundary spin-up and spin-down currents counterpropagate whereas at
the other sample boundary spin-up and spin-down currents propagate
in the same directions. Using Landauer conductance formula for the
four-terminal geometries we find that for Case \textbf{I} in
Fig.~\ref{fig:fig4}(c) with a general $N_x$, there is spin-polarized
charge current $(2N_x+1/2)Ve^2/h$ flowing from terminal 1 to
terminal 2 with a Hall voltage $V/2$, leading to a Hall conductance
$\sigma_{xy}=(4N_x+1)e^2/h$. We note that this feature differs this
state from the state of topological insulator since the Hall voltage
in the latter state is $0$\cite{Abanin2006}, not $V/2$. For Case
\textbf{II} in Fig.~\ref{fig:fig4}(d) with a general $N_x$, there is
also a spin-polarized charge current $(2N_x+1)Ve^2/h$ flowing from
terminal 1 to terminal 2 but without a Hall voltage, leading to a
resistance $1/\sigma_{xx}=h/(2N_x+1)e^2$. Interestingly, for Case
\textbf{I} in Fig.~\ref{fig:fig4}(c), there exist \emph{spin
currents} $(2N_x-3/2)Ve^2/h$ flowing from terminal 1 and $Ve^2/h$
flowing from terminal 4, as well as a \emph{spin current}
$(2N_x-1/2)Ve^2/h$ flowing to terminal 2, whereas for Case
\textbf{II} in Fig.~\ref{fig:fig4}(d), there exist a \emph{spin
current} $2N_xVe^2/h$ flowing from terminal 2 to terminal 1, as well
as a $N_x$ independent \emph{spin current} $Ve^2/h$ flowing from
terminal 3 to terminal 4. We note that a wave-vector-dependent
spin-filtering effect was also revealed recently by a calculation on
the transport problem through magnetic barriers in graphene with
Zeemann splitting\cite{DellAnna2009-2}.

A natural question is that the SMF we considered here is ideal and
changes direction abruptly, while in real conditions there exists a
length scale $l$ which is the distance covered by the SMF to change
direction. Detailed calculation shows that if $l$ is much less than
the magnetic length $l_B$, our results will be independent of $l$.
The $l$ is estimated to be $1$ nm, while for a typical magnetic
field order of $10$ T, $l_B\sim 10$ nm, satisfying the criterion.

In conclusion, graphene QHE under a modulated orbital magnetic field
has been investigated. The current-carrying edge states created by
the modulated magnetic field give rise to a novel quantized Hall
conductance, which can be checked by a standard four-terminal Hall
measurement. By varying the UMF, the four-terminal graphene sample
is expected to undergo a transition from a state with novel QHE to
one without Hall effect. The effect of Zeeman spin splitting is also
discussed and a novel state and its corresponding spin Hall currents
are predicted.

\begin{acknowledgments}
This work was supported by NSFC Projects 10504009, 10874073 and 973
Projects 2006CB921802, 2006CB601002.
\end{acknowledgments}

\end{document}